# Universal energy limits of radiation belts in planetary and brown dwarf magnetospheric systems

Drew L. Turner[1*], Savvas Raptis[1], Adnane Osmane[2], Arika Egan[1], George Clark[1], Tom Nordheim[1], Leonardo Regoli[1], and Sasha Ukhorskiy[1]

**Affiliations:**

[1]Johns Hopkins Applied Physics Laboratory, Laurel, MD, USA

[2]University of Helsinki, Helsinki, Finland

*Corresponding author. Email: Drew.Turner@jhuapl.edu

**Abstract:** Radiation belts are regions of magnetically trapped particle radiation found around all of the sufficiently magnetized planets in the Solar System and recently also observed around brown dwarfs, yet despite their ubiquity, there is not yet a general theory or model to predict the uppermost energy limits that any particular magnetospheric system's radiation belts can attain. By considering only the most fundamental loss processes, a model and corresponding theory are developed that successfully bound and explain the maximum observed energies of all documented radiation belt systems. Interestingly, this approach yields a relatively simple function for the uppermost energy limit that depends on only the surface magnetic field strength of the system. The model predicts an energy limit for *all* radiation belt systems that asymptotes at 7 ± 2 TeV (for protons and electrons), offering intriguing new insight on potential sources of galactic cosmic rays. This model is also applied to an exoplanetary system, demonstrating that the planet is likely a synchrotron emitter and showcasing the model's use for identifying candidate targets for synchrotron-emitting astrophysical systems and revealing details critical to habitability at those remote worlds.

Magnetospheric systems, such as the space environments surrounding Earth and Jupiter dominated by intrinsic planetary magnetic fields, are able to trap charged particles, concentrate them at very high intensities, and accelerate them to relativistic energies (v ~ c). Earth's radiation belts boast intense populations of trapped protons and heavy ions up to several gigaelectronvolts (GeV) [1.Mazur] and electrons up to ≥15 MeV [2.Blake]; meanwhile, Jupiter, with its strong internal dynamo and enormous magnetospheric system, retains intense populations of positive ions and electrons up to *at least* 2.5 GeV and 100 MeV, respectively [3.Kollmann; 4.de Pater and Dunn]. Every sufficiently magnetized planet in the Solar System boasts radiation belts, and radiation belts have now also been observed by imaging synchrotron emissions from a brown dwarf magnetospheric system [5.Kao]. This implies that, provided adequate conditions, radiation belts are ubiquitous in magnetospheric systems, both planetary and stellar, throughout the cosmos. Considering that, a generalized model is developed to estimate the upper energy limits on any radiation belt system. This model successfully bounds the upper energies at every observed radiation belt system throughout the Solar System and beyond and even offers predictive capability for the uppermost electron energies observed at Mercury, Earth, Jupiter, Saturn, Uranus, and Neptune. Applied to brown dwarfs, the model can explain the synchrotron emissions from brown dwarf LSRJ1835 + 3295 [5.Kao; 6.Climent] and also predict the maximum possible electron and proton energies at that system as ~7 TeV, which the model produces as a universal asymptotic limit for uppermost energy possible in planetary to brown dwarf radiation belt systems and is consistent with the unexplained break in the cosmic ray electron (and positron) spectrum around ~1 TeV [7.Aharonian; 8.Thoudam]. Furthermore, the model can be applied to exoplanetary systems, offering predictive capability to prioritize which exoplanets may be synchrotron emitters and revealing details of their potential for habitability.

Intrinsic magnetic (B) fields are generated by internal dynamo processes within large, spinning planets and stars (and even some moons), and the resulting fields tend to be dominated by a large dipole moment. In a sufficiently strong, dipole-like B-field, energetic charged particles (i.e., particles within the suprathermal to relativistic tails of collisionless plasma distributions) can be effectively trapped in the field by the Lorentz force, which results in three characteristic periodic motions [9.Northrop]: i) gyration around the field lines, ii) bounce between mirror points along the field lines, and iii) azimuthal drift perpendicular to field lines around the full system. Hamiltonian action integrals associated with each of those motions can be derived into a set of three adiabatic invariants. The invariants remain conserved so long as the underlying B-field is not changing on timescales or spatial scales comparable to the periods or scales, respectively, of the corresponding motions. The first invariant, associated with the gyromotion and magnetic rigidity, is the most difficult to invalidate, and conservation of this invariant, Mu $\propto$ K/B, where K is kinetic energy, results in betatron acceleration or deceleration provided increases or decreases, respectively, in the ambient B-field. Conservation of the second adiabatic invariant, associated with the bounce motion, results in Fermi-type acceleration or deceleration. Meanwhile, the third invariant, associated with azimuthal drift around the system, is the easiest to invalidate, resulting in radial transport within the system. Collectively, these invariants are critical to all acceleration, loss, and transport processes in radiation belt systems [10.Schulz and Lanzerotti].

A number of critical processes have been identified for radiation belt sources, acceleration, losses, and transport, particularly at Earth [11.Millan and Thorne; 12.Shprits; 13.Li and Hudson]. Here, we develop a model that is independent of any particular system's source, acceleration, or transport processes. Instead, this model focuses only on the universal limiting loss factors of trapped energetic particles in radiation belt systems. Those limiting factors include: i) the gyrosounding limit; ii) the magnetic rigidity limit; and iii) the synchrotron limit. The gyrosounding limit is the simplest, both physically and comprehensively, in that particles are assured to be lost once their energy and corresponding gyroradius are so large that the gyromotion intersects the host celestial body (i.e., the planet or star). The magnetic rigidity limit is associated with conservation of the first adiabatic invariant; when a particle's gyroradius becomes comparable to the radius of curvature of the background B-field, the first invariant is broken and the particle is randomly scattered [14.Gray and Lee]. Such scattering will ultimately result in particles rapidly "precipitating" down field lines and into the host celestial body, where they will be absorbed and lost due to collisions in dense media. The synchrotron limit is only relevant for the highest-energy, highly relativistic particles, which achieve energies where their gyromotion in the B-field results in the rapid emission of energy in the form of synchrotron radiation, ultimately limiting their maximum obtainable energy.

Our model considers the following inputs: the host celestial body's i) size (i.e., planetary or stellar radius); ii) B-field dipole moment magnitude; iii) spin rate; and iv) magnetospheric plasma density; and a range of Mu for a particular particle species (e.g., electrons, protons, alphas, etc.). Employing the approximation of a dipole field, the model is fully analytical yet only applicable in relatively close proximity to the host body (i.e., at relatively low L, where L is a magnetic field line coordinate specified as the radial distance from the center of the body to a point in the magnetic equatorial plane in units of planetary radii). The close proximity approximation is reasonable, however, considering that charged particle radiation coalesces and is most intense close to the host celestial body. With those inputs, particle energies can be derived as a function of L-shell and Mu assuming conservation of the first adiabatic invariant, as shown in the underlying color and contours in Figure 1. Here, we only show results for particles in the magnetic equatorial plane where the particles' pitch angle (i.e., angle between the particle velocity and local B-field vectors) is 90 degrees; these particles are the most stably trapped and can attain the highest energies in any dipole-like magnetospheric system. However, this model does include all stably-trapped pitch angles outside of the loss cone(s). Next, the universal limits are included as a function of location in the B-field and corresponding particle energy. The gyroradius, $\rho = \gamma m_0 v_\perp / q/B$, is calculated and converted to units of planetary radius, such that once $\rho \sim L$, we qualify a particle as having reached the gyrosounding limit and being lost from the system.

For the rigidity limit, the radius of curvature of the background B-field, $R_{cB}$, is also required, and for this, we also consider the centrifugal stretching of magnetospheric B-fields due to rapid rotation (see Methods section). In rapidly rotating systems, B-fields are pulled radially outward by centrifugal forces, resulting in a reduction of the equatorial B-field radius of curvature and the formation of "magnetodisks" at systems such as Jupiter and Saturn [15.Achilleos]. The model does not consider any additional modifications to the B-field, such as those from magnetopause or magnetotail current systems. Magnetic rigidity is a complicated factor. Based on theory [16.Speiser; 17.Sergeev; 18.Buchner and Zelenyi], simulations [19.Ashour-Abdalla], and supporting observations [20.Wilkins], a critical term discerning particles that are stably trapped, scattering chaotically, or quasi-stably trapped is $\kappa^2 = R_{cB} / \rho$. When $\rho \ll R_{cB}$ (i.e., $\kappa^2$ is very large), particles are stably trapped and fully adiabatic. When $\kappa^2 \leq 1$, particles undergo a quasi-stable bounce motion in which the trajectory is adiabatic away from the minimum in $R_{cB}$ along the field lines but goes ballistic around the minimum in $R_{cB}$ [16.Speiser; 17.Sergeev]. In a critical regime $1 < \kappa^2 < \kappa_{cr}^2$, particles behave chaotically in the field and reach a strong pitch angle diffusion limit, in which they scatter very rapidly into the loss cone of the host body (i.e., where the mirror points fall to altitudes below the collisional regime of the host and particles are lost). Based on theory and supporting simulations, $\kappa_{cr}^2$ typically ranges between 3 and 33, with an ideal value of ~8 [16.Speiser]. For our model, we consider a range of $\kappa^2$ as a function of energy and L-shell and consistent with theory in which particles are lost rapidly from a magnetospheric system.

Finally, for the synchrotron limit, the energy loss rate is calculated as $f_\rho K_{sync} \propto K^4/\rho$, where $f_\rho$ is the gyro-frequency, and when the e-folding timescale of the energy loss by synchrotron emissions becomes comparable to the relativistic drift period, $T_{drift} \propto R_p^2 qB / (\gamma m_0 L)$, where $R_p$ is planetary radius and q is a particle's charge, then particles are considered to have hit an asymptotic upper energy limit and can no longer be accelerated in the system. Figure 1 shows this model applied to protons in the HR-8799-e exoplanetary system as an example case that details each of those limits and how they appear in the [L, Mu, E] state-space.

There are seven magnetospheric systems within the Solar System: Mercury, Earth, Ganymede, Jupiter, Saturn, Uranus, and Neptune, and applying this model at each showcases its effectiveness at predicting the uppermost energy limits for radiation belt systems. For electrons, the model successfully bounds the uppermost energies that have been observed or estimated at each magnetospheric world. For protons (and heavier ions), the model also successfully bounds the observed upper energy limits at Mercury, Earth, Jupiter (including oxygen ions), Ganymede, Saturn, Uranus, and Neptune. Figure 2 shows the results for electrons and ions, respectively, at Mercury, Earth, and Jupiter (see plots for other systems in the supporting material). All of the proton and electron observations are within the upper energy limits for gyrosounding and stable trapping, respectively, as predicted by the model, and at Jupiter, the model also accurately estimates the peak of electron synchrotron emissions at L ~ 1.3. Interestingly, it seems as though the rigidity limit is most critical to electrons while the gyrosounding and/or synchrotron limits are most critical to protons and heavier ions, though ultimately, this is also dependent on the sources and source energies of particles at each system. For example, to enable electrons at Earth to get into the quasi-stable trapping regime, there would need to be either a) some sufficient source of electrons at >1 MeV at L > 10, including some inward transport process to further accelerate them, or b) direct injections on the order of 10 MeV directly and routinely into L < 6, but there are no known mechanisms that do that. However, at Ganymede, a moon-magnetosphere system embedded entirely within Jupiter's own magnetosphere, there is a sufficient source of MeV electrons from the Jovian radiation belts to provide an external supply of ~MeV electrons directly into the quasi-stable trapping regime at that system [21.Clark]. Jupiter's observed limits are ultimately bounded by the performance ranges of observatories that have probed that system, but for Neptune and Uranus and to some extent Saturn, those systems might be more limited by a lack of particle sources in the system (e.g., Neptune [22.Kollmann]) and/or additional, dominant loss processes (e.g., Saturn and its ring system [23. Roussos]). Ultimately, the observed maximum energies of radiation belt particles all fall *below* the predicted maxima from this model in all of the systems where observations are available (including *both* remote sensing *and* in situ measurements), as shown in Figure 3.

It is intriguing that, at least for electrons, each of the systems appears to reach its upper energy limit based on rigidity; that is, for all of the observed systems, electrons reach or exceed the stable trapping limit regardless of different source, acceleration, transport, and loss processes are active. This implies that regardless of either universal or distinctive source, acceleration, loss, and transport processes within any particular radiation belt system, if the source, acceleration, transport, and loss conditions are appropriate then a system can and does indeed reach those uppermost stable-trapping energy limits for electrons. Protons and ions, which in the Solar System have more abundant external sources at higher energies compared to electrons, seem more readily able to extend above the $\kappa_{cr}^2$ limit but also seem ultimately more limited by loss processes. For protons, magnetotails and cosmic rays both provide sufficient sources and direct injection of protons right into the quasi-stable trapping regime, so unlike for electrons everywhere but Ganymede, the chaotic scattering regime due to loss of rigidity is not a hard limit for protons throughout the Solar System's magnetospheres. Based on these results, this model offers distinct promise for application to astrophysical radiation belt systems elsewhere in the cosmos.

Applying the model to the ultracool brown dwarf system, LSR J1835+3259 [5.Kao], the model successfully reproduces the reported ≥15 MeV electrons synchrotron emitting at L~12 in the system. For this estimate, the observed [5.Kao] ~3 hr spin period and ~5 kG surface B-field strength were used. A stellar radius of 70,000km was assumed [5.Kao], and the plasma mass density within the system was estimated at $2 \times 10^{-21}$ kg/m$^3$, (i.e., ~1% of Jupiter's plasma density, based on [24.Saur]). With those factors, the model successfully reproduces a synchrotron emission boundary at L~11 and quasi-stable electrons (i.e., $\kappa^2 < 1$) at 15 MeV, as seen in Figure 2. Most interestingly, when other brown dwarf systems are examined with this model using a reasonable range of stellar radii (0.7 to 1.4 R$_{Jupiter}$) and surface field strengths ($10^{-2} \sim 10^0$ T), the uppermost energy limit for both electrons and protons, which is ultimately bounded by the intersection of the synchrotron and gyrosounding thresholds, asymptotes at 7 TeV, as shown in Figure 3.

These results imply that radiation belt systems from the most extreme planetary systems (Jupiter, exoplanets) and brown dwarves may contribute to the cosmic ray spectrum up to ~7 TeV, but anything above that energy limit must be accelerated in some other type of system (e.g., supernova shocks, pulsars, magnetars, active galactic nuclei, etc.). This new insight introduces many potential additional and nearby sources for the highest energy cosmic ray electrons ever observed (up to several TeV) and an explanation for the knee in the electron cosmic ray spectrum at ~1 TeV [25.Archer], a result recently supported by high-statistics measurements [7.Aharonian]. Indeed, recent results attempting to explain that knee in the electron cosmic ray spectrum [8.Thoudam], concluded that it is likely due to a sudden change in abundance of source systems, entirely consistent with the implications of this study: there are many, many more brown dwarfs in the Milky Way than there are pulsars, magnetars, supernova remnants, or other potential sources of >~TeV electrons (and positrons). The distribution of uppermost energy limits for each system sorts very nicely as a function of surface B-field strength, B$_s$. The distribution is fit very well by a functional form (shown on Figure 3) incorporating a power law at lower B$_s$ and an asymptote around 7 TeV, due to the coupled relationships between $\rho$, R$_p$, B$_s$, Mu, and L considering only the gyrosounding and synchrotron limits. Essentially, as B$_s$ increases beyond the critical limit specified by B$_0$, the intersection of the gyrosounding and synchrotron limits moves to higher L-shells because of the dependency on the drift periods; as demonstrated here for a representative range of exoplanet and brown dwarf magnetic fields (Fig. 3), this results in an asymptote for K$_{max}$ at ~7 TeV for both electrons and protons. As shown in the methods section, this relationship of K$_{max}$ as a function of B$_s$, including the power law at lower K$_{max}$ and maximum energy limit due to synchrotron energy loss, can also be derived from first principles, employing a combination of radial diffusion and synchrotron energy losses in a magnetospheric system ultimately limited by R$_p$ and B$_s$. This analytical approach also produces the same power law as the model shown in

Figure 3: β = +2.1 from the analytical and β = +2.2 ± 0.5 from the fit on Figure 3. Ultimately, this model offers both predictive capability *and* a natural maximum energy limit attainable by planetary and brown dwarf radiation belt systems.

This simple yet effective predictive model can also be applied to exoplanetary systems. Figure 1 shows the example for protons at exoplanet HR-8799-e [26.Lacour]. This gas giant planet orbits F-type star HR 8799 at an estimated distance of 16.4 AU and has a mass of 10x that of Jupiter. Assuming that it has a comparably strong surface B-field of ~40 G, a plasma density and radius comparable to Jupiter, $R_H$ ~ 1.2 $R_J$, and a spin period of around 1-hour, HR-8799-e is capable of stably trapping electrons up to 500 MeV and quasi-stably trapping protons up to 5 TeV. Most importantly from an astronomical perspective, if HR-8799-e's magnetosphere is a radiation belt system as estimated here, then it should also be emitting significant synchrotron emissions, which might render it as a particular target of interest for future observational campaigns. With the model tuned as described above, we predict peak electron synchrotron emissions around L ~ 3.6. This demonstrates how this simple model can also be used to identify potential exoplanets of interest, not only from an observational perspective [27.Lazio] but also concerning star/planet interactions [28.Ramstad and Barabash], atmospheric evolution [29.Gronoff], and habitability, since radiation levels within a planetary system are a factor of consideration for sustaining life [30.Grießmeier et al., 2015; 31.Driscoll, 2018].

**Acknowledgments:** DLT is thankful to the National Science Foundation (NSF) for funding that enabled this research.

**Funding:**

National Science Foundation grant S03464-01

**Author contributions:**

Conceptualization: DLT, SU

Methodology: DLT, SU, AO, AE, GC, SR, TN, LR

Investigation: DLT, GK, AE, TN

Visualization: DLT, SR, TN, LR

Supervision: DLT, SU

Writing – original draft: DLT

Writing – review & editing: DLT, SR, AO, AE, GC, TN, LR, SU

**Competing interests:** Authors declare that they have no competing interests.

**Data and materials availability:** All data are available in the main text or the supplementary materials.

**Methods:**

Summary of Results and Key Model Parameters
Table S1
Model Definitions and Equations
Theoretical Approach
Figure S1

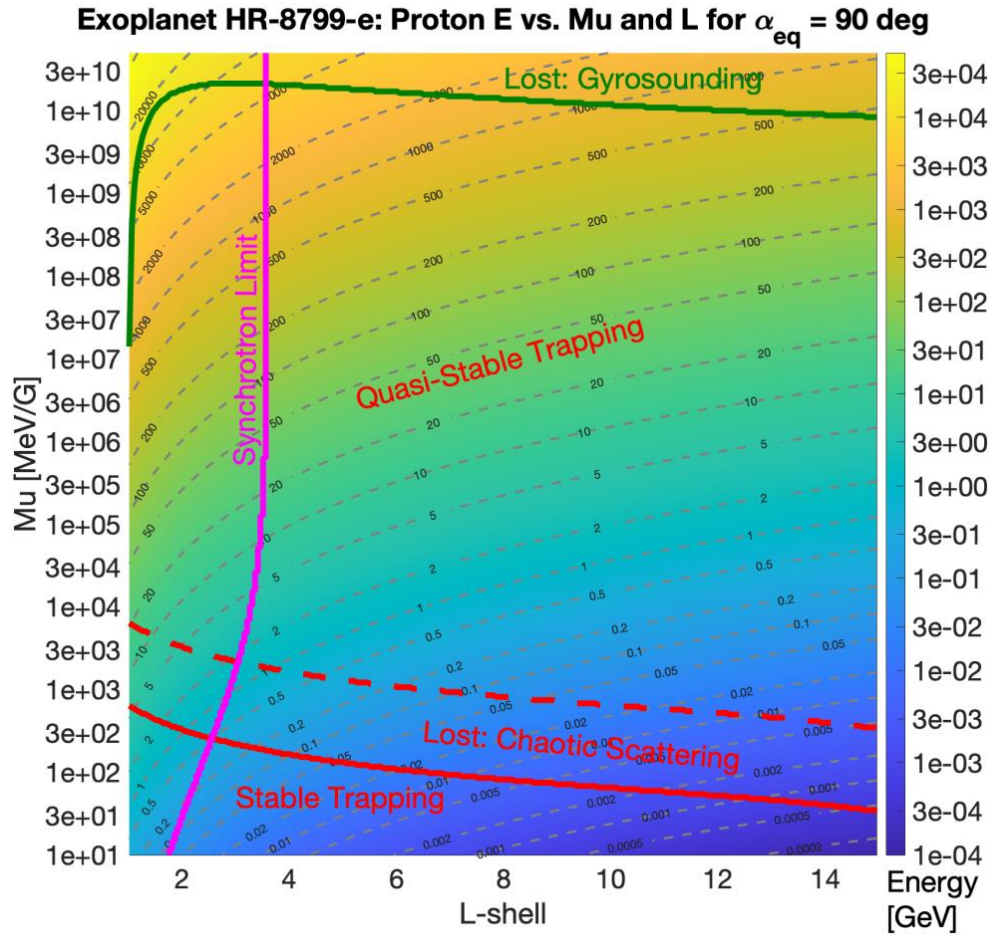

**Figure 1:** Energy limit model results applied for equatorially mirroring protons in a hypothetical radiation belt system at exoplanet HR-8799-e. Proton energy is plotted in color as a function of L-shell and first adiabatic invariant, Mu, with constant-energy contours (in units [GeV]) shown with dashed gray lines. Overplotted in this [L, Mu, E] state space are curves corresponding to the limit factors discussed in the text: i) chaotic scattering due to loss of magnetic rigidity (violation of the first adiabatic invariant) in red; ii) gyrosounding limit, in which a particle's gyroradius is as large as the L-shell resulting in sudden loss from the system, in green; and iii) the synchrotron limit, in which the energy loss timescale by synchrotron emissions is on the same order as the particle's drift period, in pink. Essentially, anything above the green line, to the left of the pink line, or between the red solid and dashed lines, should be lost very quickly from the system. However, any particles in in the regions labeled "Stable Trapping" or "Quasi-Stable Trapping" should be able to remain and accumulate within this radiation belt system. Whether a system can attain its theoretical upper energy limit ultimately depends on the different source, acceleration, transport, and loss processes active within the system, so this new model offers an estimate of the uppermost energy limits possible within a radiation belt system plus insight on whether a system might be an active emitter of synchrotron radiation.

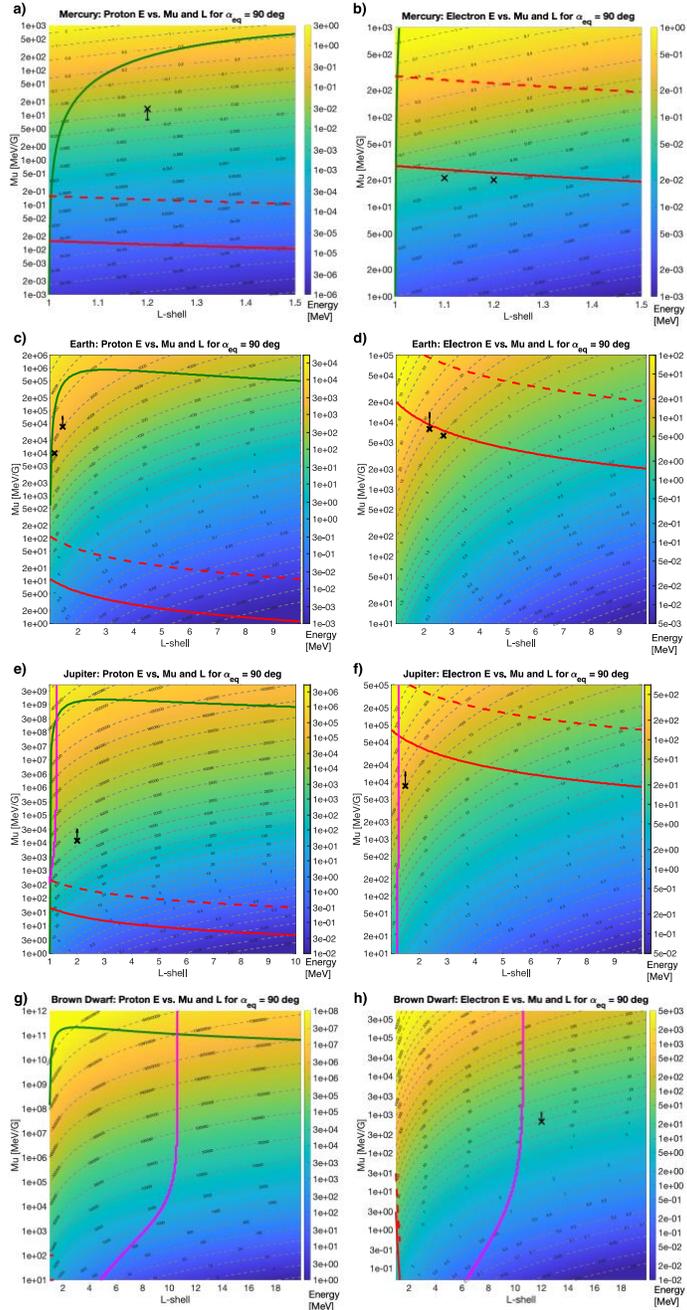

**Figure 2:** Energy limit model results applied to four different systems: Mercury, Earth, Jupiter, and brown dwarf LSR J1835+3259. Results for each plot are shown in the same format as in Figure 1. For each of the four systems, results are shown for protons in the column on the left and electrons in the column on the right. Actual observed limits are shown with black "x"s, with integral energy values shown with an upward arrow and uppermost limits shown with a downward arrow. For each system, we only show one or two of the highest energies reported at the lowest L-shells.

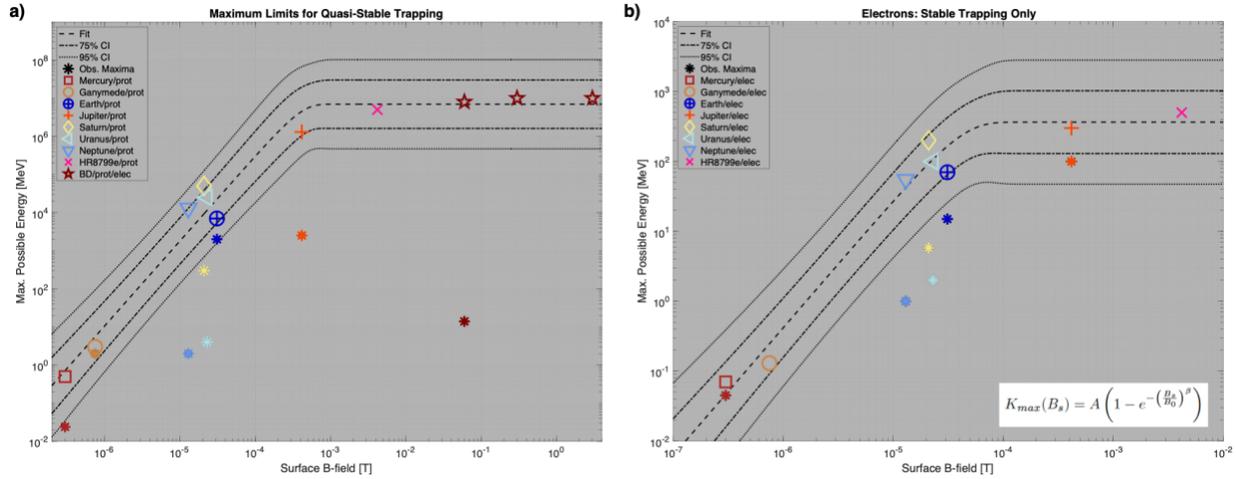

**Figure 3: a)** Modeled maximum energy limits for ions (protons for all system plus also electrons at the brown dwarf systems) for quasi-stable trapping regimes in radiation belt systems throughout the Solar System and beyond. **b)** Modeled energy limits for stably trapped electrons in various systems. Observational limits (as also noted in Figure 2) are also shown with asterisks. Maximum possible energies in each system are plotted versus surface B-field strength, which nicely organizes the data into a functional form with a power law at lower $K_{max}$ and $B_s$ and a rollover into an asymptotic limit in $K_{max}$, which is ultimately limited by the coupled intersection between the gyrosounding and synchrotron limits as a function of L and Mu in each system. The functional form for the fits is shown with the equation in the bottom right corner of panel b), with the least-squares best fits (in log space) shown with the dashed black lines and the 75% and 95% confidence intervals on each fit shown with the dash-dot and dotted lines, respectively. For the fits, the following constants are applied: ions: $A = 6.9 \pm 1.6$ TeV, $B_0 = 4.1 \pm 0.9 \times 10^{-4}$ T, $\beta = 2.2 \pm 0.5$; electrons: $A = 365 \pm 51$ MeV, $B_0 = 4.1 \pm 0.6 \times 10^{-5}$ T, $\beta = 1.8 \pm 0.3$, where A is the asymptotic limit of $K_{max}$, $B_0$ is the rollover point in $B_s$, and $\beta$ is the slope on the power law. These plots show that the uppermost, quasi-stable trapping limits successfully bound all of the radiation belt systems with observations, and for electrons at most systems (with the notable exceptions of Ganymede and the brown dwarf, as discussed in the text), the stable trapping limit appears to be the ultimate limiting threshold. This model implies that planetary and brown dwarf radiation belt systems may serve as sources of cosmic ray protons and electrons up to at most 7 TeV.

# Materials and Methods for

# Universal energy limits of radiation belts in planetary and brown dwarf magnetospheric systems

Drew L. Turner[1]*, Savvas Raptis[1], Adnane Osmane[2], Arika Egan[1], George Clark[1], Tom Nordheim[1], Leonardo Regoli[1], and Sasha Ukhorskiy[1]

[1]Johns Hopkins Applied Physics Laboratory, Laurel, MD, USA

[2]University of Helsinki, Helsinki, Finland

*Corresponding author. Email: Drew.Turner@jhuapl.edu

**The PDF file includes:**

    **Materials and Methods:**
        Summary of Results and Key Model Parameters
        Table S1
        Model Definitions and Equations
        Theoretical Approach
        Figure S1

**Materials and Methods**

**1. Summary of Results and Key Model Parameters**

Table S1 details the model parameters used for each planetary and brown dwarf magnetosphere and the corresponding maximum energies, both observed (with references for each observation listed below the table) and predicted by the new model. These results correspond to those shown in Figure 3 of the main manuscript.

**Table S1: Summary of results and key model parameters**

| Magnetospheric System | Body Radius | Dipole Strength (Bs) | Rotation Period | Plasma Density | Max Energy Observed | Max Energy Predicted |
|---|---|---|---|---|---|---|
| [] | [km] | [T] | [hours] | [#/cm^3] | [MeV] | [MeV] |
| Mercury | 2440 | 3.00E-07 | 1407.500 | 1E+02 | Elec: 4.5E-02<br>Prot: 2.4E-02 | Elec: 7E-02<br>Prot: 5E-01 |
| Earth | 6371 | 3.11E-05 | 23.934 | 1E+02 | Elec: ≥1.5E+01<br>Prot: ≥2.0E+03 | Elec: 7E+01<br>Prot: 7E+03 |
| Jupiter | 69950 | 4.17E-04 | 9.925 | 1E+02 | Elec: ≥1.0E+02<br>Prot: ≥2.5E+03 | Elec: 3E+02<br>Prot: 1E+06 |
| Ganymede | 2631 | 7.50E-07 | 171.720 | 1E+02 | Elec: ≥2.0E+00<br>Prot: ≥2.0E+00 | Elec: 1E-01<br>Prot: 3E+00 |
| Saturn | 58300 | 2.10E-05 | 10.561 | 8E+01 | Elec: 5.8E+00<br>Prot: 3.0E+02 | Elec: 2E+02<br>Prot: 5E+04 |
| Uranus | 25360 | 2.30E-05 | 17.240 | 7E-01 | Elec: 2.0E+00<br>Prot: 4.0E+00 | Elec: 1E+02<br>Prot: 3E+04 |
| Neptune | 24600 | 1.30E-05 | 16.110 | 4E-02 | Elec: 1.0E+00<br>Prot: 2.0E+00 | Elec: 6E+01<br>Prot: 1E+04 |
| HR-8799-e | 84000 | 4.20E-03 | 9.000 | 2E+03 | Unobserved | Elec: 5E+02<br>Prot: 8E+06 |
| LSRJ1835 + 3295 | 70000 | 6.00E-02 | 2.000 | 1E+00 | Elec: >14 MeV<br>Prot: Unobserved | Elec: 1E+07<br>Prot: 1E+07 |

References for observations at each system include: Mercury/protons [Walsh et al., 2013, doi:10.1002/jgra.50266]; Mercury/electrons [Ho et al., 2012, doi:10.1029/2012JA017983; Walsh et al., 2013, doi:10.1002/jgra.50266]; Earth/protons [A. Bruno*; Mazur et al., 2023, doi:10.1007/s11214-023-00962-2]; Earth/electrons [Blake et al., 1992, doi:10.1029/92gl00624; Baker et al., 2013, doi:10.1038/nature13956]; Jupiter/protons [Kollmann et al., 2021, doi:10.1029/2020JA028925]; Jupiter/electrons [de Pater and Dunn, 2003, doi:10.1016/S0019-1035(03)00068-X]; Ganymede/protons and /electrons [Kollman et al., 2022, doi:10.1029/2022GL098474]; Saturn/protons [Krupp et al., 2018, doi:10.1016/j.pss.2009.06.010; Roussos et al., 2018, doi:10.1126/science.aat1962]; Uranus/protons [Mauk, 2014, doi:10.1002/2014JA020392] and /electrons [Mauk et al., 1987, doi:10.1029/JA092iA13p15283; Mauk and Fox, 2010, doi:10.1029/2010JA015660]; Neptune/protons [Mauk, 2014, doi:10.1002/2014JA020392] and /electrons [Krimigis et al., 1990, doi:10.1029/GL017i010p01685; Mauk and Fox 2010]; brown dwarf /electrons [Kao et al., 2023, doi:10.1038/s41586-023-06138-w].

*Bruno, A., (2019), High-energy trapped protons measured by PAMELA, Reported at the IRENE Space Radiation Modeling and Data Analysis Workshop, 29-31 May 2019, Sykia, Greece

## 2. Model Definitions and Equations

The following equations were used to develop the model that generated the results presented in this study.

The relativistic first adiabatic invariant is calculated as:

$$M = \frac{K_\perp \left(K_\perp + 2m_0 c^2\right)}{2 B m_0 c^2} \quad (1)$$

where $K_\perp$ is particle kinetic energy perpendicular to B

For gyrosounding and rigidity, the following is used for the gyroradius and rigidity parameter, κ:

Gyroradius:

$$\rho = \frac{\gamma m_0 v \sin\alpha}{qB} \quad (2)$$

Rigidity parameter:

$$\kappa^2 = \frac{R_c B}{\rho} \quad (3)$$

Synchrotron energy loss rate is calculated with:

$$f_\rho K_{sync} = 88.46 \times 10^{-3} \left(\frac{K^4}{\rho}\right) \quad (4)$$

with K in [GeV] and ρ in [m], yielding energy loss per gyroperiod in [keV/$T_\rho$], where $T_\rho$ is the gyroperiod in [s].

Particle drift periods as a function of energy and L-shell are calculated with the following model described in Schulz and Lanzeroti [8.Schulz and Lanzeroti]:

$$f_D = \frac{3}{2\pi} \left(\gamma^2 - 1\right) \frac{m_0 c^2 L}{\gamma q B R_p^2} \left(\frac{D_y}{T_y}\right) \quad (5)$$

where,

$$T_0 = 1 + \frac{1}{2\sqrt{3}} \log\left(2 + \sqrt{3}\right) \quad (6)$$

$$T_1 = \frac{\pi}{6\sqrt{2}} \tag{7}$$

$$D_y = \frac{1}{12}\left(4T_0 - (3T_0 - 5T_1)\sin\alpha - (T_0 - T_1)\left(\sin\alpha \log(\sin\alpha) + \sqrt{\sin\alpha}\right)\right) \tag{8}$$

$$T_y = T_0 - \frac{1}{2}(T_0 - T_1)(\sin\alpha)^{\frac{3}{2}} \tag{9}$$

The form of the asymptotic power-law function used to fit to the max energies as a function of surface field strength, $B_S$, is:

$$K_{max}(B_s) = A\left(1 - e^{-\left(\frac{B_s}{B_0}\right)^\beta}\right) \tag{10}$$

where A defines the level of the asymptote, $B_0$ determines the turning point where the power law turns over into the asymptote, and β is the slope of the power law. That asymptotic power-law behavior is understandably physical, as derived from first principles and described below.

## 3. Theoretical Approach

We examine whether a combination of radial transport and synchrotron radiation can account for the power-law dependence of energy on surface magnetic field described in this work. We begin by estimating the minimum magnetic field strength required for synchrotron radiation to significantly impact the azimuthal drift motion of relativistic particles. It appears that that even a moderately strong magnetic field can alter the guiding center motion if we consider net damping on timescales comparable to the drift motion rather than the Larmor motion. We assume relativistic, magnetically trapped particles undergoing azimuthal drift due to the M∇B force. Next, we model radial diffusion as the mechanism responsible for radial transport and calculate the energy gain achieved before synchrotron radiation becomes dominant, limiting further energy increases. We find a power law dependence between the maximum energy and the surface magnetic field, which indicates that a combination of synchrotron radiation and radial transport is sufficient to explain the results of this study.

The damping force due to synchrotron radiation [Rybicki and Lightman, 1991, Radiative processes in astrophysics] for a particle in a constant magnetic field (assuming that the magnetic field is constant and to ignore the electric field is a first order approximation that can be shown to be valid as long as the particle loses energy faster than it can sample variation in the electromagnetic field during a drift orbit. In short, we assume that the typical fluctuations in the electric field are too small for the particles to experience significant jumps in drift shells across a few drift periods.) is given by:

$$\boldsymbol{F}_{rad} = -\frac{2}{3}\frac{r_e^2}{m_s^2 c^2}\gamma^2 \left(\frac{\boldsymbol{v}}{c} \times \boldsymbol{B}\right)^2 \frac{\boldsymbol{v}}{c}$$

in terms of the classical electron radius, $r_e$ = 2.8179x10-13 cm. Note that all quantities used thereafter are in cgs units.

The last ingredient we need is the azimuthal drift frequency $\Omega_d$, due to the M∇B drift in a dipolar field of the form $B_s R_p^3/r^3$ where $R_p$ is the planet radius, and $B_s$ is magnetic field on the surface. We ignore the curvature drift since we are doing this for 90 degrees pitch-angle particles and non-dipolar contributions that are significant in planets such as Uranus or Neptune [Stanley, 2004 ; Stanley, 2006, ]. However, a more general calculation to include any pitch angle and non-dipolar contributions can easily be provided. The drift frequency (in cgs) is given by:

$$\Omega_d = \frac{3Mc}{q\gamma L^2 R_p^2}$$

where L is the normalized radial distance: $L = r/R_p$

We can track the evolution of a collection of magnetically trapped particles in a dipolar field with fixed adiabatic invariant M in terms of the distribution function f = f(L, φ, t; M) radial transport kinetic equation:

$$\frac{\partial f}{\partial t} + \Omega_d \frac{\partial f}{\partial \varphi} = \frac{\partial (\dot{M} f)}{\partial M}$$

where the change in the first adiabatic invariant dM/dt (M-dot) is due to synchrotron losses. We write the synchrotron term on the right-hand side, because radiation, like collisions, can violate conservation of phase-space density. If we decompose the distribution function in terms of a drift average part $f_0 = f_0(L, t)$ that is independent of the azimuthal dependence φ, and a fast part that tracks the evolution of the distribution function on timescales less than the drift period, δf = δf (L, φ, t; M), write the latter in terms of Fourier modes $\delta f = \Sigma_m \delta f_m e^{\{im\varphi\}}$, we find the following equation:

$$\frac{\partial \delta f_m}{\partial t} + im\Omega_d \delta f_m = -\frac{\partial \dot{M}}{\partial M}\delta f_m$$
$$= -\left(2^{3/2}\frac{r_e^2}{m_s^2 c^2}B^{5/2}M^{1/2}m_s^{1/2}\right)\delta f_m$$

where we used the equations for M and synchrotron radiation force in the ultra-relativistic regime ($p_{perp}/m_s c \gg 1$; note: we also performed the calculation for the case when $K/m_s c^2 \sim 1$ and where the ratio of v/c in the radiation force equation need to be included, and not simply equal to 1 as you would for the ultra-relativistic regime, and we found the exact same result as below) to write the rate of change of M with respect to time, i.e. dM/dt ~ 2M dp$_{perp}$/dt / dp$_{perp}$. The equation immediately above has a very simple solution with an oscillating part that accounts for the drift motion, and a decaying part that arises due to synchrotron damping:

$$\frac{1}{\delta f_m}\frac{\partial \delta f_m}{\partial t} = -im\Omega_d - \left(2^{3/2}\frac{r_e^2}{m_s^2 c^2}B^{5/2}M^{1/2}m_s^{1/2}\right) \Rightarrow \delta f_m = \delta f_m(t=0)e^{-im\Omega_d t - \nu t}$$

where the damping due to synchrotron radiation is given by:

$$\nu = \left(2^{3/2}\frac{r_e^2}{m_s^2 c^2}B^{5/2}M^{1/2}m_s^{1/2}\right)$$

Physically, the decaying synchrotron portion of the equation is telling us that magnetic trapping at a fixed drift shell will cease when $\nu\Omega_d \geq 1$ in less than a drift period. If particles come off their drift shells in less than one drift period, then there is no possibility for them to experience radial transport anymore on the initial drift shell. We can estimate the magnitude of the magnetic field required for this effect to take place. Equivalently, we bypass this argument with the kinetic equation and compute the following non-dimensional quantity and determine when it compares to unity:

$$\left|\frac{\dot{M}}{M}\frac{1}{\Omega_d}\right| \geq 1$$

Computing this equation, we find that the minimum magnetic field (in cgs) for synchrotron damping to nullify magnetic trapping is given by:

$$\boxed{B_{max} \geq \left(\frac{9}{8}\frac{m_s^2 c^4}{q r_e^2 R_p^2}\right)^{1/3}\frac{1}{L^{2/3}}.}$$

When we set L=1, we find magnetic field values (after conversion from Gauss to Tesla) consistent with the plateau transition in Figure 3, i.e. B$_s$ ≥ 8x10$^{-5}$ T for electrons at Jupiter and B$_s$ ≥ 1x10$^{-2}$ T for protons at Jupiter. The maximum magnetic field is a function of the particle mass, so the higher the particle mass the larger the magnetic field limit. But the value B$_{max}$ for electrons do not vary significantly for Saturn or Earth either with a dependence on planetary radius that goes as R$_p$$^{-2/3}$, and the threshold we found in Figure 3 is well matched by the equation for B$_{max}$ here.

What is the maximum energy a particle can gain when being magnetically trapped? If the first adiabatic invariant from injection is preserved before going through synchrotron damping at L = 1, then maximum kinetic energy K$_{max}$ is the geometric mean of its rest mass and the product of M and B$_{max}$(L=1):





$$K_{max} \simeq (2MB_{max}(L=1))m_s c^2)^{1/2}$$

Note that the above analysis also tells us that each drift shell has an associated maximum energy. So this analysis indicate that if particles are carried through any form of betraton towards the inner belts, and B < $B_{max}$ they can have their maximum kinetic energy grow as $K_{max} \sim B_s^{1/2}$ across the radiation belts. However, if B ~ $B_{max}$, the particles stop experiencing drift motion and never reach higher energy than the one given by the maximum energy equation immediately above.

To explain the power law dependence in the fitting function for Figure 3, we first expect that the constant $B_0$ should be proportional to $B_{max}$ found in the previous section, since the maximum energy plateau when $B_s \geq B_{max}$ and when $B_s/B_0 \ll 1$ then $K_{max} \sim (B_s/B_0)^\beta$. Also, based on the estimate for the damping due to synchrotron radiation, $K_{max}$ is certainly a function of the surface magnetic field $B_S$ but also of L, so perhaps a good approach to seek the above power law is by solving the distribution of particles experiencing radial transport and then integrating over all drift shells to find the average $K_{max}$.

We first assume that the particles are experiencing radial transport due to some electrostatic fluctuations that violate the third adiabatic invariant. We can of course consider more complex radial transport drivers, but we are trying to figure out if we can find a parametric dependence of $K_{max}$ that is consistent with the fitting equation in Figure 3. The simplest statistical mode we know of is quasi-linear radial diffusion so we will use that, but we should keep in mind that as long as the radial transport equation can be expressed in terms of a Fokker-Planck equation, with or without higher order effects, that our analysis most likely holds (or put differently, transport does not have to be quasi-linear).

The quasi-linear radial diffusion equation for electrostatic fluctuation with electric field **E** = δE(φ) **φ**|$_{unit}$ is given by Falthammar [1965, doi:10.1029/JZ070i011p02503] and Osmane [2024, doi:10.48550/arXiv.2409.12649]:

$$\frac{\partial f_0}{\partial t} = L^2 \frac{\partial}{\partial L}\left(\frac{D_{LL}}{L^2}\frac{\partial f_0}{\partial L}\right)$$

where the radial diffusion coefficient, $D_{LL}$, is written as:

$$D_{LL} = c^2 \frac{|\delta K_{\varphi,m}|^2}{B_S^2 R_P^2} L^4$$

in terms of the electric field power spectrum |δE(φ, m)|², the surface magnetic field $B_s$ and the planetary radius $R_p$. Note that the diffusion coefficient has units of s$^{-1}$. The radial diffusion equation combined with





a spatially dependent radial diffusion equation can be transformed to the classical 1D diffusion equation. We first normalize time:

$$\tau = c^2 \frac{|\delta K_{\varphi,m}|^2}{B_S^2 R_P^2} t$$

and write the equation in terms of the inverse of the normalised radial distance x=1/L. Doing so results in the following diffusion equation:

$$\frac{\partial f_0}{\partial \tau} = \frac{\partial^2 f_0}{\partial x^2},$$

This can be solved analytically in several ways. The solution to the radial diffusion equation is given by:

$$\begin{aligned} f_0 &= \frac{n_s}{2\sqrt{\pi}} \frac{\exp(-x^2/4\tau)}{\sqrt{\tau}} \\ &= \frac{n_s}{2\sqrt{\pi}} \frac{B_S R_P}{c|\delta K_{\varphi,m}|} \frac{1}{\sqrt{t}} \exp\left(-\frac{1}{4} \frac{L^2}{D_{LL}} \frac{1}{t}\right) \end{aligned}$$

for t > 0 and with $n_s$ a normalization constant that gives the number of particle per unit volume, which we assumed to be conserved to find the above solution. Here, we need to make an additional modification to the solution (i.e., equation immediately above) to make sure that whenever particles reach the maximum energy, at any L value where the synchrotron radiation is too strong for a full drift orbit to be completed, that they are stopped moving inward radially. This is the equivalent to introducing a barrier in the phase-space density. It should be stressed that we are still assuming that particles are magnetically trapped, but that since they experience synchrotron damping, their drift period goes down (note that the synchrotron radiation will also impact the guiding center position and we can compute this effect which will appear as first and second order effects in the kinetic equation. But it seems to me that synchrotron radiation becomes undeniably significant when the drift period, which is the dominant component of the guiding center trajectories, is altered). However, one also needs to make sure that the solution with the barrier satisfies the radial diffusion equation for all times. Such a solution is given by using the method of images and results in the following solution:

$$f_0 = \frac{n_s}{2\sqrt{\pi}} \frac{\exp(-x^2/4\tau)}{\sqrt{\tau}} - \frac{n_s}{2\sqrt{\pi}} \frac{\exp\left[-(x-2b)^2/4\tau\right]}{\sqrt{\tau}},$$

where b = $(B_{max}(L=1) / B_s)^{3/7}$, the value of x when $B(x) = B_{max}(x)$. Note that $f_0 = 0$ for x=b results in the synchrotron radiation ceasing further betatron energization and any further corresponding inward radial transport.

With the underlying assumption that radial transport is responsible for energization, $K_{max}$ is given by:





$$K_{max}/m_s c^2 \simeq \left(\frac{2MB_{max}}{m_s c^2}\right)^{1/2}$$

$$= \left(\frac{2MB_{max}(L=1)}{m_s c^2}\right)^{1/2} \frac{1}{L^{1/3}}$$

Let's compute the average value of $K_{max}$ when radial diffusion is taking place, and thus when the energy is weighted by the solution of $f_0$ (above equation):

$$\left\langle \frac{K_{max}}{m_s c^2} \right\rangle = \int_{x=0}^{x=b} dx \left(\frac{2MB_{max}(L=1)}{m_s c^2}\right)^{1/2} x^{1/3} \frac{n_s}{2\sqrt{\pi}} \left[\frac{\exp(-x^2/4\tau)}{\sqrt{\tau}} - \frac{\exp\left[-(x-2b)^2/4\tau\right]}{\sqrt{\tau}}\right]$$

$$= \frac{n_s}{2\sqrt{\pi}} \left(\frac{2MB_{max}(L=1)}{m_s c^2}\right)^{1/2} \int_{x=0}^{x=b} dx\, \delta \frac{x^{1/3}}{\sqrt{\tau_{max}}} \left[\exp\left(-\frac{\delta^2 x^2}{4\tau_{max}}\right) - \exp\left(-\frac{\delta^2 (x-2b)^2}{4\tau_{max}}\right)\right]$$

where we renormalized $\tau$, to make apparent the dependence on the surface magnetic field $B_s$, with the following quantities: $\tau_{max} = c^2\, |\delta K_{\varphi,m}|^2 / (B_{max}\, R_p)^2\, t$, $\delta = B_s / B_{max}(L=1)$, and $b = \delta^{-3/7}$. Thus, what we define above as some average $K_{max}$ is a function of the surface magnetic field $B_S$. When solved numerically, we find that the energy dependence on $B_s$ is consistent with a power-law, up to the point where $B_s / B_{max} \sim 1$, as shown in Figure S1 below:

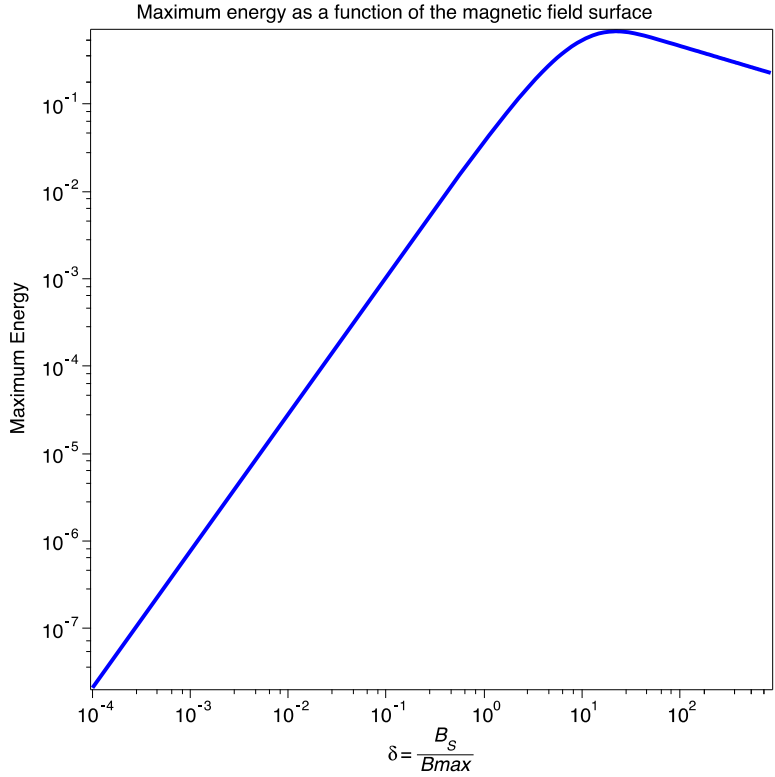

**Figure S1:** Solution to the equation for Kmax immediately above to show the dependence of the maximum kinetic energy versus surface magnetic field. The dependence of the logarithm of K is linearly proportional to the logarithm of Bs, but when Bs > Bmax, energization rolls over and stops.

Note that the dip down beyond the peak in $B_s$ / $B_{max}$ is simply because this approach considers L = 1 as the location where $B_{max}$ is defined in the example shown here. For systems in which $B_{max}$ occurs at higher L-shells (i.e., $B_{max}$ occurs at L > 1, beyond the surface, in systems where $B_s$ is larger than that estimated here), the turnover will occur at higher δ, which is entirely consistent with the results in Figure 3 that combine a number of different systems and ultimately reveal an asymptotic value in $K_{max}$. The point is that regardless of where, in L-shell for L > 1, $B_{max}$ is reached in this theory, $K_{max}$ will be constant at the level shown here, *and* ultimately, δ(L = 1) is the most extreme case on the low $B_s$ side for any synchrotron emitting system. This is all entirely consistent with the complementary approach described in the study (i.e., with results shown in Figure 3), including the power law slope on the lower δ values: β = +2.1 from this analytical approach and β = +2.2 ± 0.5. from the fit to results in Figure 3, from which we also establish a prediction for the asymptotic $K_{max}$ at 7 TeV.